\documentclass[11pt,a4paper]{article}

\usepackage{graphicx}
\usepackage{a4wide}
\usepackage{amsfonts}
\usepackage{amsmath}
\advance \textheight by 1mm

\usepackage{amsthm}
\newtheorem{lemma}{Lemma}
\newtheorem{theorem}{Theorem}
\newcommand{\area}{\mathop{\mathrm{area}}}
\begin{document}
\title{Strictly Convex Drawings of Planar Graphs}
\author{Imre B\'ar\'any\thanks{R\'enyi Institute, Hungarian Academy of
    Sciences, PoB 127, Budapest 1364, Hungary, barany@renyi.hu, and
    Mathematics, University College London, Gower Street, London WC1E
    6BT, UK.}
%Partially supported by Hungarian National Science Foundation Grants No. T 037846 and T 046264.}
 \and G\"unter Rote\thanks{Freie Universit\"at Berlin,
Institut f\"ur Informatik, 
Taku\-stra\ss e~9, 14195 Berlin, Germany,
rote@inf.fu-berlin.de}}
\maketitle

\begin{abstract}
  Every three-connected planar graph with $n$ vertices has a drawing on an
  $O(n^{2}) \times O(n^{2})$ grid in which all faces are strictly convex
  polygons.  These drawings are obtained by perturbing (not strictly) convex
  drawings on $O(n) \times O(n)$ grids.  Tighter bounds are obtained when the
  faces have fewer sides.
  In the proof, we derive an explicit lower bound on the
  number of primitive vectors in a triangle.
\end{abstract}

% CG, DM
% ACM: F.2.2; G.2.2

\newif\ifnarrow
%\narrowtrue % some figures have to be scaled down
%auto-ignore
\section{Introduction}
A \emph{strictly convex} drawing of a planar graph is a drawing with
straight edges in which all faces, including the outer face, are
strictly convex polygons, i.~e., polygons whose interior angles are
less than $180^\circ$.

\begin{theorem} \label{main}
\begin{enumerate}
\item[\rm(i)]
 A three-connected planar graph with $n$ vertices
in which every face has at most $k$ edges
has a {strictly convex} drawing
 on an $O(nw) \times O(n^2k/w)$ grid of area $O(n^3k)$,
for any choice of a parameter $w$ in the range $1\le w \le k$.
\item[\rm(ii)]
 In particular, every
 three-connected planar graph with
  $n$ vertices has a {strictly convex} drawing on an $O(n^2) \times
  O(n^2)$ grid, and on an $O(n) \times O(n^3)$ grid.
\item[\rm(iii)]
For $k\le 4$,
 an $O(n) \times O(n)$ grid suffices.
\end{enumerate}
The drawings can be constructed in linear time.
\end{theorem}

When referring to a $W\times H$ grid of width $W$ and height $H$,
the constant hidden in the $O$-notation is on the order of 100 for the
width
and on the order of 10000 for the height.
This is far too much for applications where
one wants to draw graphs on a computer screen, for example.
For the case $w=1$, the bound is tighter: the grid size is approximately
$14n \times 30n^2k$.
 For part~(iii) of the theorem, the grid size is at most
$14n \times 14n$, and if the outer face is a triangle, it is
$2n \times 2n$.

%\pagebreak

The main idea of the proof is to start with a (non-strictly) convex embedding,
in which angles of $180^\circ$ are allowed,
and to
perturb the vertices to obtain strict convexity.  We will use an embedding
with special properties that is provided by the so-called \emph{Schnyder embeddings},
which are introduced in Section~\ref{sec:schnyder}.

\paragraph{Historic context.}
The problem of drawing graphs with straight lines has a long history.
It is related to realizing three-connected planar graphs as
three-dimensional polyhedra.  By a suitable projection on a plane, one
obtains from a polyhedron a straight-line drawing, a so-called
\emph{Schlegel diagram}.  The faces in such a drawing are
automatically strictly convex.
By a projective transformation, it can be arranged that
the projection along a coordinate axis is possible, and hence a suitable realization as a grid
polytope gives rise to a grid drawing of the graph.
However, the problem of realizing a
graph as a polytope is more restricted: not every drawing with
strictly convex faces is the projection of a polytope.
In fact, there is an exponential gap between the
known grid size for strictly convex planar drawings and
for polytopes in space.

The approaches for realizing a graph as a polytope or for drawing it
in the plane come in several flavors.  The classical methods of
Steinitz (for polytopes) and F\'ary and Wagner (for graphs) work incrementally,
making local modifications to the graph and adapting the geometric
structure accordingly.  Tutte~\cite{t-crg-60,t-hdg-63} gave a
``one-shot'' approach for drawing graphs that sets up a system of
equations.  This method yields also a polytope via the Maxwell-Cremona
correspondence, see~\cite{richter-gebert-book}.
All these methods give embeddings that can be drawn on an integer grid but
require an exponential grid size (or even larger, if one is not careful).

The first methods for straight-line drawings of graphs on an
$O(n)\times O(n)$ grid were proposed for triangulated graphs,
independently by de Fraysseix, Pach and Pollack~\cite{fpp-hdpgg-90}
and by Schnyder~\cite{s-epgg-90}.  The method of de Fraysseix, Pach
and Pollack~\cite{fpp-hdpgg-90} is incremental: it inserts vertices in
a special order, and modifies a partial grid drawing to accommodate
new vertices.  In contrast, Schnyder's method is another ``one-shot''
method: it constructs some combinatorial structure in the graph, from
which the coordinates of the embedding can be readily determined
afterwards.  Both methods work in linear time.  $O(n)\times O(n)$ is
still the best known asymptotic bound on the size of planar grid drawings.

If graphs are not triangulated, the first challenge is to get faces which
are convex.
(Without the
convexity requirement one can just add edges until the graph becomes
triangulated, draw the triangulated
supergraph and remove the extra edges from the drawing.)  Many
algorithms are now known that construct convex
 (but not necessarily strictly convex) drawings
with
$O(n)\times O(n)$ size, for example by Chrobak and
Kant~\cite{ck-cgd3c-97}
%, title =       "Convex grid drawings of 3-connected planar graphs"
% and Chrobak, Goodrich and Tamassia~\cite{cgt-cdgtt-96}
(\`a~la Fraysseix, Pach and Pollack);
or
Schnyder and Trotter~\cite{st-ce3cp-92}
and Felsner~\cite{f-cdpgo-01}, see also \cite{bfm-cd3cp-05}
(\`a la Schnyder).
Our algorithm builds on the output of Felsner's algorithm,
which is described in the next section.
Luckily, this embedding has some special features, which our algorithm uses.

The idea of getting a strictly convex drawing by perturbing a convex
drawing was pioneered by Chrobak, Goodrich and
Tamassia~\cite{cgt-cdgtt-96}.  They claimed to construct strictly
convex embeddings on an $O(n^3) \times O(n^3)$ grid, without giving
full details, however.
This was improved to
 $O(n^{7/3}) \times O(n^{7/3})$ in \cite{r-scdpg-05}.
In this paper we further improve the ``fine perturbation'' step of \cite{r-scdpg-05}
to obtain a bound of
 $O(n^2) \times O(n^2)$ for grid drawings. Theorem~\ref{main} gives better bounds when
the faces have few sides,
and we allow grids of different aspect ratios (keeping the same total area).

In the course of the proof, we need explicit (not just asymptotic) lower
bounds on the number of primitive vectors in certain triangles.
A primitive vector is an integer vector which is not a multiple of another
integer vector; hence, primitive vectors can be used to characterize the
directions of polygon edges. The existence of many short primitive vectors
is the key to constructing strictly convex polygons with many sides.
 These lower
bounds are derived in Section~\ref{proof-grid}, based on elementary
techniques from the geometry of numbers.

\section{Preliminaries: Schnyder Embeddings of Three-Connected Plane Graphs}
\label{sec:schnyder}

Felsner~\cite{f-cdpgo-01} (see also \cite{f-gedp-03,bfm-cd3cp-05}) has
extended the straight-line drawing algorithm of Schnyder, which works
for triangulated planar graphs, to arbitrary three-connected graphs.
It constructs a drawing with very special properties, beyond just
having convex faces. These properties will be crucial for the
perturbation step.

\begin{figure*}[htb]
  \begin{center}
\noindent
\ifnarrow
    \includegraphics[width=\textwidth]{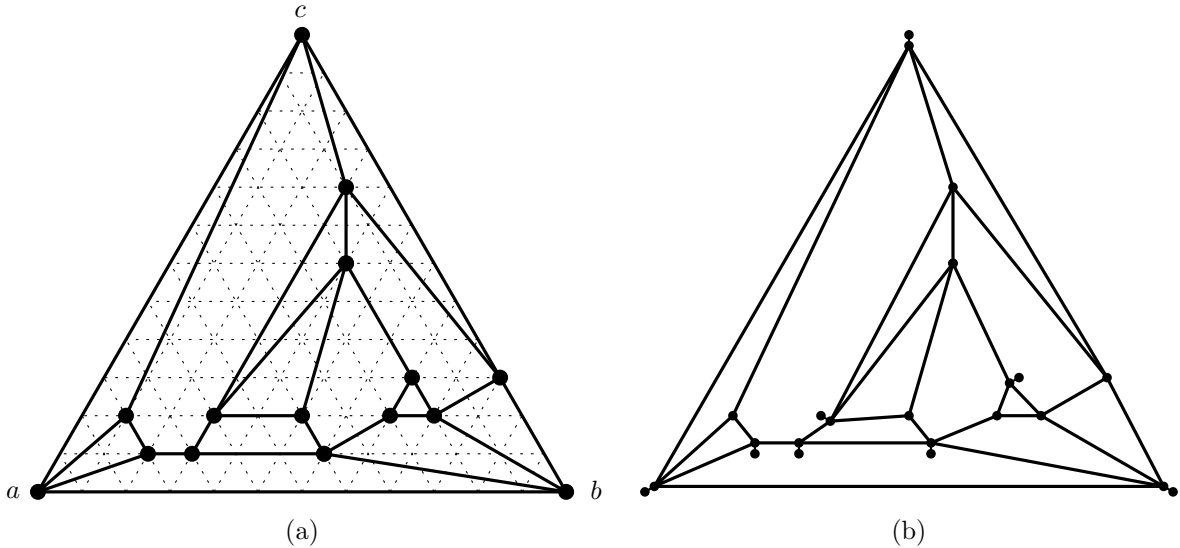}
\else
    \includegraphics{beispiel-and-perturb-a}
\fi
    \caption{(a) A Schnyder embedding on a hexagonal grid
and (b) on the refined grid after the initial (rough) perturbation}
    \label{fig:example-perturb}
    \label{fig:example}
  \end{center}
\end{figure*}

Felsner's algorithm works roughly as follows.
The edges of the graph are covered by three directed trees which are rooted at
three selected vertices $a$, $b$, $c$ on the boundary,
forming a \emph{Schnyder wood}.
The three trees define for each vertex $v$ three paths from $v$ to the
respective root, which partition the
graph into three regions. Counting the faces in each region gives three
numbers $x,y,z$ which can be used
% (after normalizing them
as barycentric coordinates for the point $v$ with respect to the points $a$,
$b$, and $c$.
Selecting $abc$ as an equilateral triangle of side length $f-1$ (the number of
interior faces of the graph) yields vertices which lie on a hexagonal grid
formed by equilateral triangles of side length~1, see
Figure~\ref{fig:example}a. Since $f\le 2n$ this yields a drawing on a grid of
size $2n\times 2n$.

\begin{figure}[htb]
  \begin{center}
    \includegraphics{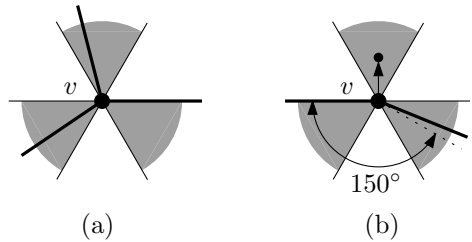}
    \caption{(a) Each closed shaded wedge contains exactly one edge incident
      to $v$. There may be additional edges in the interior of the white sectors.
(b) A typical situation at a vertex which is perturbed.}
    \label{fig:wedges}

  \end{center}
\end{figure}

This straight-line embedding has the following important property
(see \cite[Lemma~4 and Figure~11]{f-cdpgo-01}, \cite[Fact~5]{bfm-cd3cp-05}):
\begin{quote}
  \emph{The Three Wedges Property.}  Every vertex except the corners
  $a,b,c$ has exactly one incident edge in each of the three closed
  $60^\circ$ wedges shown in Figure~\ref{fig:wedges}a.
\end{quote}
From this it follows immediately that there can be no angle larger
than $180^\circ$, and hence all faces are convex. Moreover, it follows
that the interior faces $F$ have the \emph{Enclosing Triangle Property}, see
Figure~\ref{fig:perturb-one-polygon}a (\cite[proof of
Lemma~7]{f-cdpgo-01}, \cite[Lemma~2]{bfm-cd3cp-05}):
\begin{quote}
\emph{The Enclosing Triangle Property}.
  Consider the line $x=\textrm{const}$ through the point of~$F$ with maximum
  $x$-coordinate, and similarly for the other three coordinate
  directions.  These three lines form a triangle $T_F$ which encloses
  $F$.  Then all vertices of $F$ lie on the boundary of $T_F$, but $F$
  contains none of the vertices of $T_F$.
\end{quote}

It follows that interior faces with $k\le 4$ sides are already strictly convex.
Throughout, we will call $T_F$ the \emph{enclosing triangle} of the face~$F$.

The Schnyder wood and the coordinates of the points can be calculated in
linear time.
Recently, Bonichon, Felsner, and Mosbah~\cite{bfm-cd3cp-05}, have improved the
grid size to $(n-2)\times (n-2)$. However, the resulting drawing does not have
the Three Wedges Property.
An alternative algorithm for producing an embedding with a property similarly
to the Enclosing Triangle Property is sketched in Chrobak, Goodrich and
Tamassia~\cite{cgt-cdgtt-96}.  It proceeds incrementally in the spirit of the
algorithm of de~Fraysseix, Pach and Pollack~\cite{fpp-hdpgg-90} and takes
linear time.  From the details given in~\cite{cgt-cdgtt-96} it is not clear
whether the embedding has also the Three Wedges Property, which we
need for our algorithm.
The original algorithm of Chrobak and
Kant~\cite{ck-cgd3c-97} achieves a weak form
of the Three Wedges Property, where $F$ is permitted to contain vertices of
$T_F$. Maybe, this algorithm can be modified to obtain
 the Three Wedges Property, at the expense of a constant-factor blow-up in the grid size.

\begin{figure}[htb]
  \begin{center}
\ifnarrow
    \includegraphics[scale=0.95]{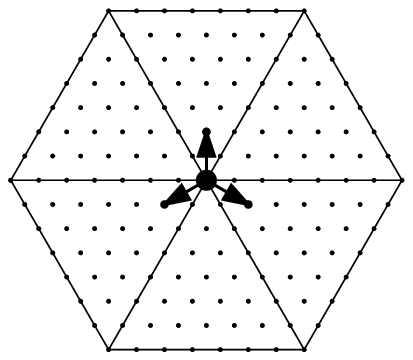}
\else
    \includegraphics{perturb-one-point}
\fi
    \caption{The three possible new positions for a single vertex in the rough
      perturbation.
(Only the three boundary vertices $a$, $b$, $c$ are pushed in directions
opposite to these.)}
    \label{fig:perturb-one-point}
  \end{center}
\end{figure}

\begin{figure*}[htb]
  \begin{center}
\ifnarrow
    \includegraphics[width=0.95\textwidth]{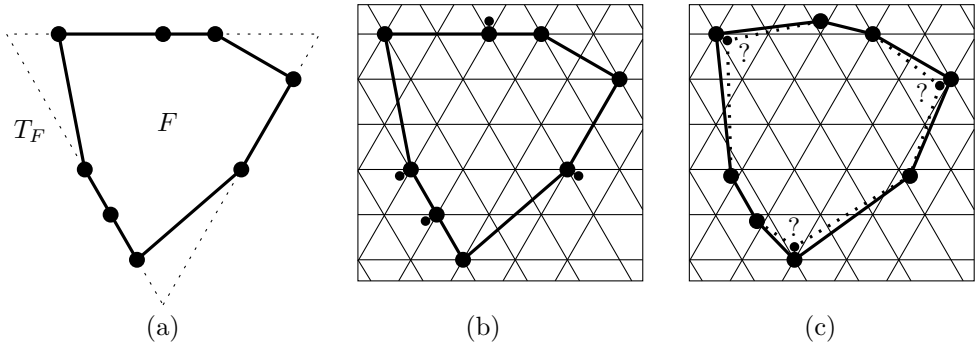}
\else
    \includegraphics{perturb-one-polygon}
\fi
    \caption{(a) A typical face $F$ constructed by the convex embedding algorithm.
(b) The new positions of the vertices of $F$ which are pushed out
are indicated. (c) The result of the rough perturbation.
      The perturbation of the vertices with question marks
      depends on the other faces incident to these vertices.}
    \label{fig:perturb-one-polygon}
  \end{center}
\end{figure*}

\section{Rough Perturbation}
\label{sec:initial-perturbation}
Before making all faces strictly convex, we perform an initial perturbation on
a refined grid which is smaller by only a constant factor.
This preparatory step will ensure that the subsequent ``fine perturbation'' can
treat each face independently.

\begin{figure}[htb]
  \begin{center}
\ifnarrow
    \includegraphics[scale=0.9]{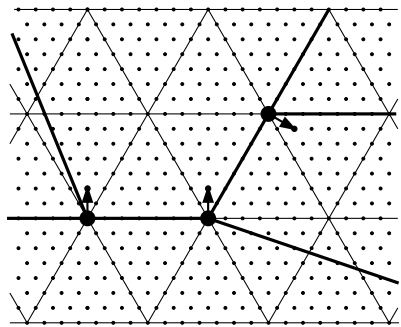}
\else
    \includegraphics{gitter7}
\fi
    \caption{Example of the rough perturbation.}
    \label{fig:perturb-small-example}
  \end{center}
\end{figure}

We overlay a triangular grid which is scaled by a factor of $1/7$, see
Figures~\ref{fig:perturb-one-point}
and~\ref{fig:perturb-small-example}.
A point may be moved to one of the three possible positions
shown in Figure~\ref{fig:perturb-one-point}, by a distance of $\sqrt3/7$.
The precise rules are as follows:
A vertex $v$ on an interior face $F$ is moved if and only if
the following two conditions hold.
\begin{enumerate}
\item [(i)] The interior angle of $F$ at $v$ is larger than
  $150^\circ$ (including the possibility of a straight angle of
  $180^\circ)$; and
\item [(ii)] $v$ is incident to an edge of $F$ which lies on the enclosing triangle~$T_F$.
\end{enumerate}
See Figure~\ref{fig:wedges}b for a typical case.  Such a vertex is
then pushed ``out'', perpendicular to the edge of $T_F$.  We call the
angle between the two edges incident to $F$ and $v$ the \emph{critical
  angle} of~$v$.  For a boundary vertex different from $a,b,c$, the
exterior angle is the critical angle, but these vertices are not
subject to the rough perturbation.  The three corners $a$, $b$, and
$c$ are treated specially: they are pushed straight \emph{into} the triangle by
the rough perturbation, as illustrated in Figure~\ref{fig:example}.
Examples can be seen in Figure~\ref{fig:perturb-one-polygon}b--c and
Figure~\ref{fig:perturb-small-example}.  The result of perturbing the
example in
Figure~\ref{fig:example}a is shown in Figure~\ref{fig:example-perturb}b.

There can be no conflict in applying the rules 
by regarding a vertex $v$ as
part of different faces: the bound of $150^\circ$ on the angle, together with
the Three Wedges Property ensures that
there is at most one critical angle for every vertex
(Figure~\ref{fig:wedges}b).

\begin{figure}[tb]
  \centering
\ifnarrow
  \includegraphics[width=\textwidth]{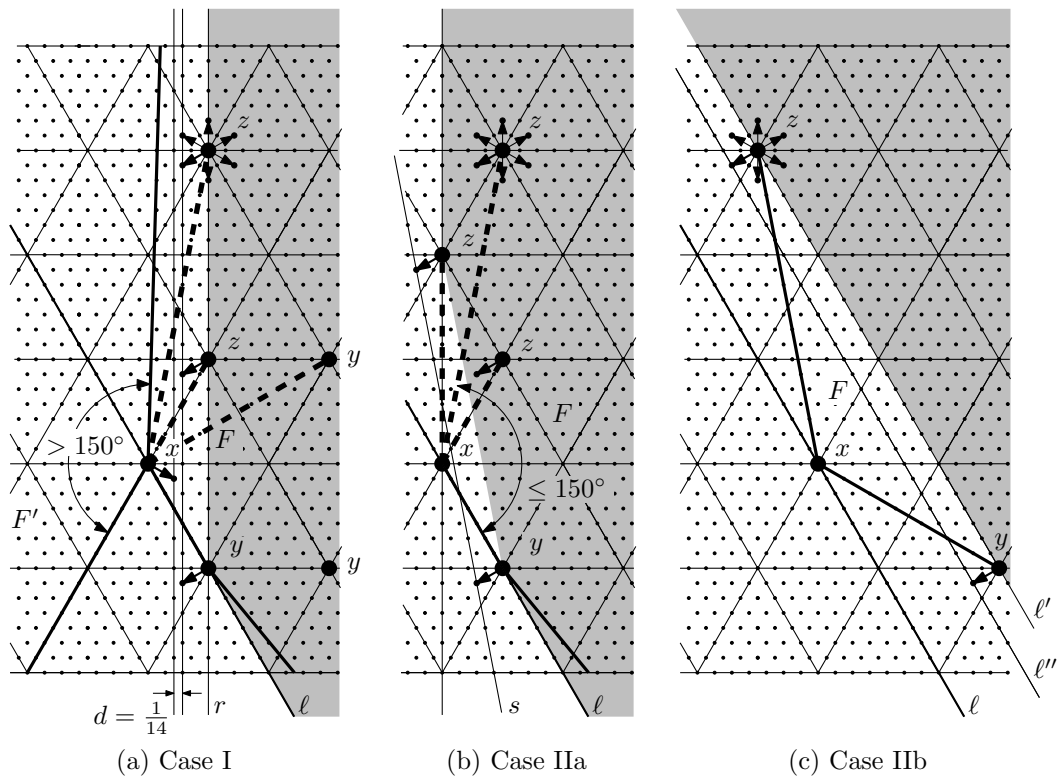}
\else
  \includegraphics{lemma1-proof}
\fi
  \caption{The cases in the proof of Lemma~\ref{small-perturb}. The figures
    show possible locations for the neighbors $y$ and $z$ of~$x$.}
  \label{fig:proof-lemma-1}
\end{figure}

The result has the following properties:
\begin{lemma}\label{small-perturb}
%  \begin{enumerate}
%  \item 
After the rough perturbation, all faces are still convex.

%\item 
Moreover, 
if each vertex is additionally perturbed within a disk of radius
 $1/30$, the only concave angle that might arise at a
  vertex $v$ is the critical angle of $v$.
%% 1/28 is the tight bound
%  \end{enumerate}
\end{lemma}

\begin{proof}
  It is evident that no critical angle can become bigger than~$180^\circ$.
  For non-critical angles, this is also easy to see
  (cf.~Figure~\ref{fig:perturb-one-polygon}c).
  (In fact, the second statement is a strengthening of this claim.)

  We now prove this second statement of the lemma
% (which is a strengthening of the first statement),
  by considering different cases.
  The reader who is satisfied with the existence of \emph{some} small
  enough perturbation bound $\varepsilon > 0$ may skip the rest of the
  proof. We continue to show that we can choose $\varepsilon=1/30$.

  Consider a non-critical
  angle $yxz$ at a vertex $x$ in a face $F$.  We assume without loss of
  generality that $x$ lies on the \emph{lower left} edge $\ell$ of the enclosing
  triangle~$T_F$.
  
  Case I. The point $x$ is incident to a critical angle of another face $F'$,
  and thus $x$ is pushed out of $F'$. \\
Without loss of
  generality,
we can assume that $x$ 
lies on the lower right edge of $T_{F'}$, and thus $x$
is
  perturbed in the lower right direction, as in
  Figure~\ref{fig:proof-lemma-1}a. 
(The other case, when $x$ 
lies on the upper edge of $T_{F'}$ and is pushed vertically upward, is symmetric.)
By the definition of critical angles, the
  angle in $F'$ must be bigger than $150^\circ$. This excludes from $F$ all points
  vertically above $x$ or to the left of $x$.  The upper neighbor $z$
  of $x$, which is a grid point, is therefore restricted to a closed halfplane
  right of a vertical line $r$ at distance $1/2$ from $x$.  The lower neighbor
  $y$ must lie on or above the line $\ell$ that bounds the enclosing
  triangle~$T_F$.  Thus, $y$ and $z$
  are restricted to the shaded area in
  Figure~\ref{fig:proof-lemma-1}a.
  Even if all three points are perturbed by the rough perturbation, they are
  still separated by a vertical strip of width $d=\frac12 - 2\cdot \frac 3{14} =
  \frac 1{14}$. An additional perturbation of $\frac 1{30}<\frac 1{2\cdot 14}$
  cannot make the angle at~$x$ larger than $180^{\circ}$.
  
  Case II. The point $x$ not perturbed by the initial perturbation.

  Case IIa.
  The point $x$ has a neighbor on $\ell$.\\ We can assume w.l.o.g.\ that it is the lower
  neighbor $y$, see
  Figure~\ref{fig:proof-lemma-1}b.
  The angle $yxz$ must be at most $150^\circ$ because otherwise
  $x$ would be critical. It means that $z$ cannot lie to the left of~$x$, and
  thus $y$ and $z$ are restricted to the shaded area in
  Figure~\ref{fig:proof-lemma-1}b.
  Even if they are perturbed, they remain above the line $s$, which is
  obtained by offsetting the edge of the shaded region that is closest to~$x$.
  The distance from
  $x$ to $s$ is $ 1/7\cdot\sqrt{3/7} \approx 0.0935 > \frac 2{30}$.
%
% $s$ goes through (-3/14, 13/7*sqrt{3/4}) and (4/14, -8/7*sqrt{3/4}).
% equation of $s$: 3*sqrt{3} x + y = 2*sqrt{3}/7
% distance of unperturbed line 3*sqrt{3} x + y = sqrt{3} is 1/2*sqrt{3/7}
% ==> grid of "1/7" reduces distance by 1/7 * 5/2*sqrt{3/7}
% ==> grid of "1/k" reduces distance by 1/k * 5/2*sqrt{3/7}
% ==> remaining distance is (1-5/k) * 1/2*sqrt{3/7}
  Thus, there is enough space to additionally perturb the points $x$, $y$ and
  $z$ without creating a concave angle. (Actually, the vertex $x$ will
  {not} even be perturbed in the fine perturbation.)
  
  Case IIb.  The point $x$ has no neighbors on $\ell$, see
  Figure~\ref{fig:proof-lemma-1}c.\\ This means that $y$ and $z$
 lie on or beyond the next grid line $\ell'$ parallel to $\ell$.
 The
  rough perturbation can move them closer to $\ell$, but they remain beyond
  another parallel line $\ell''$
whose distance from $x$ is $
  5/7\cdot\sqrt{3/4}\approx0.618$. This leaves plenty of space for additional
  perturbations of $x$, $y$, and $z$.
\end{proof}

After the rough perturbation, we will subject every vertex $v$ that is
incident to a critical angle to an additional small perturbation of a distance
at most 1/30.  The lemma ensures that, in order to achieve convexity at $v$
without destroying convexity at another place, we only have to take care of
\emph{one} incident face when we decide the final perturbation of $v$.  We can
thus work on each face independently to make it strictly convex.

\begin{figure}[htb]
  \centering
  \includegraphics{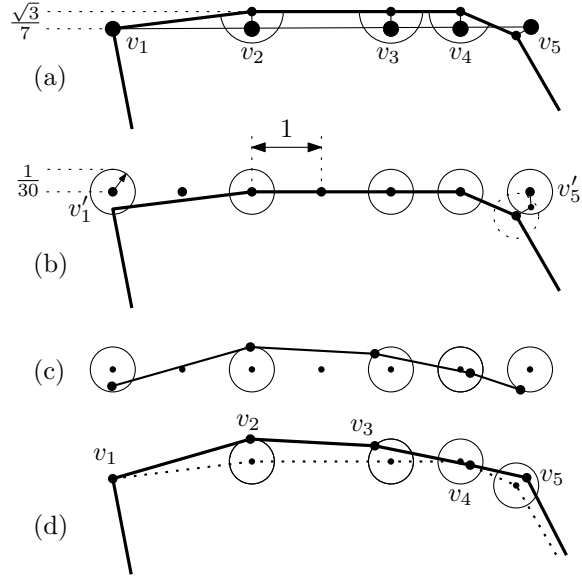}  

  \caption{The setting of the fine perturbation process: (a) The
    initial situation after the rough perturbation.
The angles in which it is necessary to ensure a convex angle are marked.
(b) The circles in which the fine perturbation is performed. The size of
the circles is exaggerated to make the perturbation more conspicuous.
(c) A strictly convex polygon inside the circles.
(d) The final result.
}
  \label{fig:fine-perturb-overview}
\end{figure}

\section{Fine Perturbation}
\label{sec:perturbation}
We will now discuss how we go about achieving strict convexity of all faces.
The rough perturbation helps us to reduce this
task to the case of regularly spaced points on a line (Section~\ref{sec:setting}).  In
Section~\ref{convex-chains-section}, we will describe in detail how the
perturbed strictly convex chain is constructed for this special case.

\subsection{The Setting after the Rough Perturbation}
\label{sec:setting}

After the rough perturbation, we are in the following situation.  Consider a
maximal chain $v_2,v_3,\ldots,v_{K-1}$ of successive critical angles on a
face~$F$.  These angles must be made strictly convex by perturbing them inside
their little disks. (The two extreme angles at $v_2$ and $v_{K-1}$ might
already be convex.)
The vertices $v_2,v_3,\ldots,v_{K-1}$
 lie originally on a common edge of the enclosing triangle $T_F$,
We first discuss the case when the vertices lie on the upper
edge~$\ell$ of $T_F$, forming a horizontal chain, as in
Figure~\ref{fig:fine-perturb-overview}a.  (The extension to the other two
cases is discussed in Section~\ref {diagonal}.)
  According to
Lemma~\ref{small-perturb} we have to ensure that these critical angles are
smaller than $180^\circ$ after the perturbation.  In
Figure~\ref{fig:fine-perturb-overview}a, these are the vertices $v_2$, $v_3$,
and $v_4$. Let us call these vertices \emph{critical vertices}.  In addition,
we look at the two adjacent vertices $v_1$ and $v_K$ on~$F$. 
 By the choice of a maximal chain, they are not critical for~$F$.
 They may lie on
the same line as the critical vertices, as the vertices $v_1$ and $v_5$ in
Figure~\ref{fig:fine-perturb-overview}a, or they might lie below this line.
%
%For the sake of the analysis,
To guide the perturbation of the points $v_2,\ldots,v_{K-1}$,
we pretend that $v_1$ and $v_K$ are part of the chain, and
we create \emph{surrogate positions} $v_1'$ and $v_K'$ for these neighbors:
First we move them from their original positions vertically upward to $\ell$; if they don't land on a grid
point, we move them outward by 1/2 unit.  
Since the angles at $v_2$ and $v_{K-1}$ are bigger than $150^\circ$,
we are sure that $v_1',v_2,\ldots,v_{K-1},v_K'$ lie on $\ell$ in this order.
Finally, we subject $v_1'$ and $v_K'$ to the same
rough perturbation as the critical vertices between them,
and move them vertically upward.

We place a disk of radius $1/30$ around every perturbed point on this edge,
including the two surrogate positions,
see Figure~\ref{fig:fine-perturb-overview}b.
%The effect of the creation of surrogate positions is that % the centers of
%these circles
%lie in a row.
% are placed as if all points were perturbed (from their original
% positions) in the same direction as the critical vertices, so that they
% form a regular row of circles.
%
%This will permit a more uniform treatment 
In the next step, to be described in Section~\ref{convex-chains-section},
we find a
strictly convex chain which selects one vertex out of each little disk, as
shown in Figure~\ref{fig:fine-perturb-overview}c.

This will make all angles at $v_2,\ldots,v_{K-1}$ strictly convex.
Finally, we use these perturbed positions for our critical vertices,
but for $v_1$ and $v_K$, we ignore their
perturbed surrogate positions, see
Figure~\ref{fig:fine-perturb-overview}d.  
% for intermediate points which were inserted, and also
%The position of $v_1$ and $v_K$ is determined independently:
The true position of $v_1$ or $v_K$ may be determined by a different face in which
it forms a critical angle (as is the case for $v_5$ in the example),
or it might just keep its original position (like $v_1$ in the example).
We only have to check that the angle at the left-most and right-most
critical vertex ($v_2$ and $v_4$ in this case) remains convex:
\begin{lemma}
  Replacing the perturbed surrogate position $v_1'$ and $v_K'$ of the points
  $v_1$ and $v_K$ by their true positions does not destroy convexity at
  their neighbors $v_2$ and $v_{K-1}$ in~$F$.
\end{lemma}
\begin{proof}
  We first show that the rough perturbation does not actually perturb $v_1$
  and $v_K$ to their surrogate positions $v_1'$ or $v_K'$. It is conceivable
  that, say, $v_1$ lies on $\ell$ and is perturbed upwards because of its
  critical angle in a different face $F'$, see Figure~\ref{fig:contra}.
  However, this would contradict the Three Wedges Property for $v_1$ and~$F$, creating two
  incident edges in a sector in which only a unique incident edge can exist.
\begin{figure}[htbp]
  \begin{center}
    \includegraphics{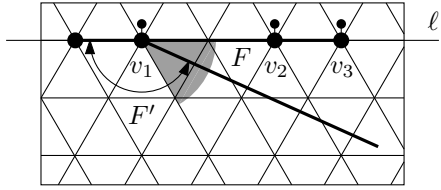}  
    \caption{A neighbor of a critical vertex cannot be perturbed in the same direction.}
    \label{fig:contra}
  \end{center}
\end{figure}
  
  Thus we conclude that $v_1$ and $v_K$ lie below or on $\ell$,
  and they are either perturbed not at all or in a direction
  below~$\ell$.
  
  Vertices $v_2$ and $v_4$ in the example of
  Figure~\ref{fig:fine-perturb-overview} represent the possible extreme cases
  that have to be considered.
  $v_5$ represents a vertex that is pushed downward in the rough perturbation,
  and then subjected to a fine perturbation anywhere in its little circle.
  For visual clarity, the circles in
  Figure~\ref{fig:fine-perturb-overview} have been drawn with a much larger
  radius than $1/30$. Since the circles are actually small enough, the angle
  at $v_4$ will be convex no matter where the point $v_5$ is placed in its own
  circle.  (This position is determined when the critical face of~$v_5$ is
  considered.)
A similar statement holds at $v_2$, where the perturbed surrogate position of $v_1$ in
Figure~\ref{fig:fine-perturb-overview}c is replaced by the \emph{original} position of
$v_1$; this will always turn the edge $v_2v_1$ counterclockwise and thus
preserve convexity at $v_2$.

The argument works also for a chain of vertices on an exterior edge of
the enclosing triangle. In this case,
$v_2,v_3,\ldots,v_{K-1}$ are perturbed around their original position
on $\ell$,
whereas the neighbors $v_1$ and $v_K$ are moved inside the triangle
and below~$\ell$.
Geometrically, the situation looks similar as for vertex $v_1$ in
Figure~\ref{fig:fine-perturb-overview}, except that $v_1$ is not pushed
down straight but at a $-30^\circ$ angle.
This movement is large enough to ensure convexity at~$v_2$.
\end{proof}

\subsection{Convex Chains in the Grid}
\label{convex-chains-section}

We have a number $K$ of vertices $0=a_1<a_2<\cdots <a_K\le 2n-1$ on a
horizontal line
which form part of an array of $2n$ consecutive grid points.  We want to
\begin{figure*}[t]
  \begin{center}
\ifnarrow
    \includegraphics[scale=1]{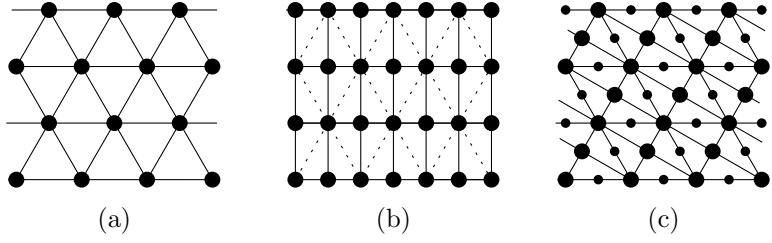}
\else
    \includegraphics{square-grid2}
\fi
    \caption{The hexagonal grid (a) is contained in a rectangular grid (b).
A hexagonal grid twice refined (c) contains rectangular grids in three
different directions. One of these rectangular grids is highlighted by thicker points.
}
    \label{fig:square-grid}
  \end{center}
\end{figure*}
perturb them into convex position.
If the faces of the embedding have at most $k$ sides, then $K\le k$.
It is more convenient to work
with a rectangular grid.  So we extend the hexagonal grid to a
rectangular grid as shown in Figure~\ref{fig:square-grid}.  This
grid will be refined sufficiently in order to allow a strictly convex
chain to be drawn inside a sequence of circles.  Figure~\ref{fig:grid}
gives a schematic picture of the situation.  (This drawing is not to
scale.) 
It is more convenient to discuss the construction of an \emph{upward} convex
chain.  Inside each disk (of radius
$1/30$) we fit a square of side length $1/50$, which is subdivided
into a subgrid of width $w$ and height $h$. More precisely, we are
looking for a sequence of points $p_i=(x_i,y_i)$ in these circles,
whose coordinates
measure the distance from the lower left corner of the first circle in units of little
grid cells.
Two successive circle centers at distance $1$ in terms of the original grid
have a distance of $S:=50w$ when measured in subgrid units.
Thus we are looking for integer coordinates
that satisfy $a_i\cdot S \le x_i \le  a_i\cdot S + w$ and $0\le y_i \le h$.
Eventually, when the whole subgrid is scaled
to the standard grid $\mathbb{Z} \times \mathbb{Z}$, $x_i$ and $y_i$ will
become true distances again. The total size of the resulting integer grid will
be $O(nw)\times O(nh)$.

\begin{figure*}[htb]
  \begin{center}
\ifnarrow
  \includegraphics[width=\textwidth]{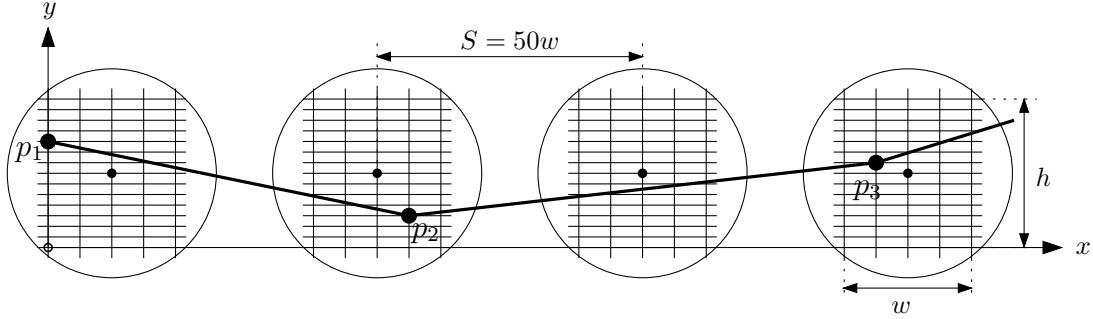}
\else
  \includegraphics{grid-example}
\fi
    \caption{A convex chain formed by grid points in the circles.
(Again, the radius of the circles is drawn much too large
      compared to their distance.)}    \label{fig:grid}
  \end{center}
\end{figure*}

The convex chain $p_1,p_{2},\ldots,p_K$ has a descending part up to a
point with minimum $y$-coordinate and an ascending part.
We choose the two points with minimum $y$-coordinate to lie in the middle:
We define $M :={\lfloor K/2\rfloor+1}$ and set
$y_{M-1}=y_{M}=0$.
We will only describe the construction of the ascending chain from
$p_{M}$ to the right. The left half is constructed symmetrically.

The direction between two grid points is uniquely specified by a
\emph{primitive vector}, a vector whose components are relatively prime.  We
now take a sequence of primitive vectors $q_1,q_2,\ldots,q_{K-M}$, $q_i=
%\binom
({u_i},{v_i})$ with $0<u_i\le w$ and $v_i>0$, in order of increasing slope
$v_i/u_i$.  Then we choose the difference vectors $\Delta p$ as appropriate
multiples of these vectors, in the following way.  We have already defined
$y_M:=0$, and we choose $x_M$ arbitrarily within the permitted range of
$x$-coordinates.  Having defined $p_{M+i-1}$, we define
$$
p_{M+i} := p_{M+i-1} + s\cdot q_i
$$
by adding as many copies of $q_i$ as are necessary to bring
$x_{M+i}$ into the desired box:
$$a_{M+i}\cdot S\le x_{M+i} \le  a_{M+i}\cdot S+ w$$
Since this box has width $w$, and $u_i\le w$, this is always possible.

We need $K-M\le K/2$ primitive vectors $q_i$ (including the vector
$(1,0)$ %\binom10$
 from
$p_{M-1}$ to $p_{M}$.)
The following theorem ensures that we can find these vectors
in a triangle of sufficiently large area.

\begin{theorem}\label{grid-in-triangle}
The right triangle $T = (0,0), (w,0), (w,t)$,
where $w\ge 1$, $w$ integer, and $t\ge 2$,
contains at least $wt/4$ primitive vectors.
\end{theorem}

The general proof is given in Section~\ref{proof-grid}.
We can however easily give an explicit solution for the special case $t=2$
(corresponding to the choice $w=k$ below, which leads to the most balanced
grid dimensions):
In this case, we can simply take % the vector $(1,0)$ and 
the $1+\lfloor w/2 \rfloor$ vectors $(w,1)$, $(w-1,1)$, \dots, $(\lceil w/2
\rceil,1)$. %, in order of increasing slope.

We use Theorem~\ref{grid-in-triangle} as follows.
We choose an arbitrary width $w\le k$ for the boxes.
By Theorem~\ref{grid-in-triangle}, we can set $t := \max\{2,2K/w\}$
to ensure that we find at least $K/2$ primitive vectors in the triangle~$T$.
The slope of these vectors is bounded by $t/w$.
Let us estimate the necessary height~$h$ of the boxes.
The last point $p_K$ is connected to $p_M$ by a chain of vectors with slope
at most $t/w$. The distance of $x$-coordinates is at most the width of
the whole grid on which the graph is embedded, i.~e., at most $S\cdot 2n=O(wn)$;
hence the difference in $y$-coordinates is at most $t/w\cdot O(wn) =
O(tn)=O(kn/w)$.
It follows that the height $h$ of the boxes is $O(kn/w)$.
The total height of the resulting grid is $O(hn)=O(kn^2/w)$.

This leads to part (i) of Theorem~\ref{main}.
Part (ii) is an easy corollary.
As an extreme case, we can set $w=1$ and perform only vertical
perturbations. We get $h\le 2kn$ (without any additional constants
depending on~$S$).

\begin{figure*}[htb]
  \begin{center}
\ifnarrow
    \includegraphics[scale=0.83]{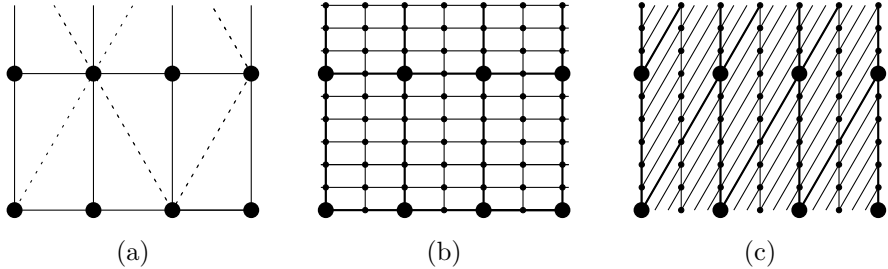}
\else
  \includegraphics{sheared-grid}
\fi
    \caption{A rectangular grid~(a),
its $2\times 6$ refinement~(b),
 and a shearing~(c) of
the refined grid. Its grid-points coincide with the untransformed grid.}
    \label{fig:sheared-grid}
  \end{center}
\end{figure*}

\subsection{Perturbation of Vertices on Diagonal Lines}
\label{diagonal}

So far, we have treated only a sequence of vertices on a horizontal straight
line. The same scheme can be applied to lines of the two other directions by
applying the shearing transformation
$\binom xy \mapsto \binom x{y+\sqrt3/2\cdot x}$
or
$\binom xy \mapsto \binom x{y-\sqrt3/2\cdot x}$
which moves points only in vertical direction.
% to the subgrid.
If $h$ is a multiple of $w$, the transformation will produce a grid like in
Figure~\ref{fig:sheared-grid}c which is contained in the original grid of
Figure~\ref{fig:sheared-grid}b.
For the range of parameters which is interesting for the theorem
($w\le k$), the height $h$ of the subgrid is never smaller than the
width $w$; thus, the choice of $h$ as a multiple of $w$ does not
change the asymptotic analysis.
One needs to reduce the size of the little square subgrid to ensure that the
sheared square still fits inside the circle, and one has to adjust the
quantity $S$ accordingly.  In addition, we have to select $h$ and $w$ as
multiples of 14, to accommodate the grid of the rough perturbation and the
refined rectangular grid of Figure~\ref{fig:square-grid}b.  All of this
changes the analysis only by a constant factor.

For the case of a uniform stretching of both dimensions ($w=h$), one
referee has pointed out a simpler alternative method.
After a blow-up by a factor of two, the original triangular grid contains
rectangular grids in all three grid directions,
Figure~\ref{fig:square-grid}c. Two further refinements by the factor
7 (for the rough perturbation) and then by the factor
$w$ are sufficient to accommodate the fine perturbation.
\ifnarrow\else\par

\fi
On the exterior edges, the points must of course be perturbed to form an
\emph{outward} convex chain.

% For part (iii) of the theorem we just observe that
% when the faces are at most pentagons, the first part of the algorithm
% will
% produce faces with at most $K=4$ vertices in a row.
% It is then easy to perturb them into strictly convex position, using
% just two possible positions in each little disk.
% The two middle vertices remain unperturbed, and the two extreme
% vertices are perturbed upward.

For part (iii) of the theorem we have already mentioned that interior faces
with $k\le 4$ sides are already strictly convex.  If the outer face has 4
edges, it contains a single vertex on one of the sides of the outer triangle.
The rough perturbation is thus sufficient to make the outer face strictly convex.

The whole procedure, as described above, is quite explicit and can be
carried out with a linear number of arithmetic operations. We calculate the
$O(k)$ primitive vectors $q_i$ only once and
store them in an array.  Then, for every actual sequence of vertices on
an edge, we can construct the perturbation very easily.
The primitive vectors in the triangle $(0,0)$, $(w,0)$, $(w,t)$ according
to Theorem~\ref{grid-in-triangle} can be selected from the $O(wt)=O(k)$ grid
points in linear time with a sieve method.

\subsection{Numerical Experiments}
\label{sec:open}

\begin{table}[tb]
  \centering
\noindent\vbox{\offinterlineskip
  \def\comma{,}
  \catcode`,=\active
  \def,{{\comma}}
\halign{\vrule\ \ \strut \hfil $#$ \
 &\vrule\ \ \hfil $#$ \ &\ \hfil $#$ \ \ \
 &\vrule\ \ \hfil $#$ \ &\ \hfil $#$ \ \ \ \vrule
\cr
\noalign{\hrule}
&\span \textrm{optimal}\hfil\hfil
&\span \textrm{greedy}\hfil
%\vrule
\cr
w=h&n&\omit $(w+1)/n$  &n&\omit $(w+1)/n$ \vrule\cr
\noalign{\hrule}
          0  &      2 &  0.5000  &      2 &  0.5000  \cr
          1  &      4 &  0.5000  &      4 &  0.5000  \cr
          2  &       6 &  0.5000  &       6 &  0.5000  \cr
          4  &      10 &  0.5000  &       8 &  0.6250  \cr
         6  &      14 &  0.5000  &      12 &  0.5833  \cr
         8  &      16 &  0.5625  &      14 &  0.6429  \cr
         10  &      20 &  0.5500  &      18 &  0.6111  \cr
        12  &      22 &  0.5909  &      18 &  0.7222  \cr
%        14  &      24 &  0.6250  &      20 &  0.7500  \cr
%        16  &      28 &  0.6071  &      24 &  0.7083  \cr
%        18  &      30 &  0.6333  &      26 &  0.7308  \cr
         20  &      32 &  0.6562  &      28 &  0.7500  \cr
         40  &      58 &  0.7069  &      48 &  0.8542  \cr
%        60  &      82 &  0.7439  &      64 &  0.9531  \cr
%        80  &     104 &  0.7788  &      82 &  0.9878  \cr
        100  &     122 &  0.8279  &      96 &  1.0521  \cr
%       120  &     144 &  0.8403  &     112 &  1.0804  \cr
%       140  &     160 &  0.8812  &     124 &  1.1371  \cr
%       160  &     178 &  0.9045  &     138 &  1.1667  \cr
%       180  &     196 &  0.9235  &     150 &  1.2067  \cr
        200  &     212 &  0.9481  &     164 &  1.2256  \cr
        400  &     366 &  1.0956  &     276 &  1.4529  \cr
%       600  &     508 &  1.1831  &     380 &  1.5816  \cr
%       800  &     638 &  1.2555  &     472 &  1.6970  \cr
      1,000  &     758 &  1.3206  &     562 &  1.7811  \cr
%     1,200  &     870 &  1.3805  &     644 &  1.8649  \cr
%     1,400  &     982 &  1.4267  &     724 &  1.9351  \cr
%     1,600  &   1,094 &  1.4634  &     806 &  1.9864  \cr
%     1,800  &   1,196 &  1.5059  &     878 &  2.0513  \cr
      2,000  &   1,292 &  1.5488  &     948 &  2.1108  \cr
      4,000  &   2,206 &  1.8137  &   1,610 &  2.4851  \cr
%     6,000  &   3,030 &  1.9805  &   2,200 &  2.7277  \cr
%     8,000  &   3,758 &  2.1291  &   2,726 &  2.9351  \cr
     10,000  &   4,468 &  2.2384  &   3,230 &  3.0963  \cr
%    12,000  &   5,140 &  2.3348  &   3,716 &  3.2295  \cr
%    14,000  &         &          &   4,170 &  3.3576  \cr
%    16,000  &         &          &   4,622 &  3.4619  \cr
%    18,000  &         &          &   5,056 &  3.5603  \cr
     20,000  &   7,592 &  2.6345  &   5,472 &  3.6552  \cr
     40,000  &         &          &   9,250 &  4.3244  \cr
%    60,000  &         &          &  12,562 &  4.7764  \cr
%    80,000  &         &          &  15,620 &  5.1217  \cr
    100,000  &         &          &  18,484 &  5.4101  \cr
%   120,000  &         &          &  21,222 &  5.6546  \cr
%   140,000  &         &          &  23,834 &  5.8740  \cr
%   160,000  &         &          &  26,376 &  6.0662  \cr
%   180,000  &         &          &  28,824 &  6.2448  \cr
    200,000  &         &          &  31,192 &  6.4119  \cr
    400,000  &         &          &  52,626 &  7.6008  \cr
%   600,000  &         &          &  71,446 &  8.3980  \cr
%   800,000  &         &          &  88,772 &  9.0119  \cr
  1,000,000  &         &          & 105,012 &  9.5227  \cr
% 1,200,000  &         &          & 120,468 &  9.9612  \cr
% 1,400,000  &         &          & 135,298 & 10.3475  \cr
% 1,600,000  &         &          & 149,634 & 10.6928  \cr
% 1,800,000  &         &          & 163,504 & 11.0089  \cr
  2,000,000  &         &          & 177,046 & 11.2965  \cr
  4,000,000  &         &          & 299,494 & 13.3559  \cr
\noalign{\hrule}
}
}

  \caption{The length of the longest strictly convex $n$-gon in a
    sequence of square cells of
    size $w\times w$, regularly spaced at distance $S=50w$.}
  \label{tab:exp}
\end{table}

We have presented a general systematic solution for finding a convex
chain by selecting grid-points from a sequence of boxes.  One can find
the \emph{optimal} (i.e., longest) convex chain in polynomial time by dynamic
programming,
as described in more detail below.
Results of some experiments are shown in the first
column of Table~\ref{tab:exp}.  We restrict ourselves to the standard situation of
selecting an $n$-gon from $n$ adjacent boxes ($K=n$) which are squares
($w=h$).  For several different sizes $w$, we computed the largest $n$
such that a strictly convex $n$-gon can be found in a sequence of
cells of size $w\times w$.  The factor $(w+1)/n$ determines the
necessary grid size $w$ in terms of~$n$.  (By the convention of
Figure~\ref{fig:grid}, a ``$w\times w$'' grid consists of $(w+1)^2$
vertices; thus we give the fraction $(w+1)/n$ instead of $w/n$.)
Since the convex chain consists of a monotone decreasing and a
monotone increasing part, connected by a horizontal segment in the
middle,
the necessary height $w+1$ is at least $0.5\,n$.
We see that this trivial lower bound is achieved for small values of~$n$.
The factor $(w+1)/n$ increases with $n$, but not very fast.  (The
rectangular $w\times h$ boxes constructed in the proof of
Theorem~\ref{main} would have $w/n=1$, but $h/n=100$.)

The dynamic programming algorithm computes, for each point $p$ in the
$w\times h$ box, and for each possible previous point $p'$ in the
adjacent box to the left, the longest ascending and strictly convex
chain (of length $i$) for which $p_{i-1}=p'$ and $p_{i}=p$.  Knowing
$p'$ and $p$, it can be determined which points in the next box are candidate
endpoints $p_{i+1}$ of a chain of length $i+1$. One can argue that,
among these points $p_{i+1}$ that are reachable as a continuation of $p'p$,
only the $w+1$ lowest points on each vertical line are candidates for endpoints $p_{i+1}$ that form part of an
optimal chain. Theoretically, the complexity of this algorithm is
therefore $O(w^3h^2)$. %, where $\Delta y$.
It turns out that, with few exceptions, every point $p$ has only one
predecessor point $p'$ that must be considered: all other predecessor
points $p_{i-1}$ have either a larger slope of the vector $p-p_{i-1}$
or they are reached by a shorter chain.  Therefore, the algorithm
runs in $O(w^2h)=O(wkn)$ time, in practice.

 A simple greedy approach for selecting the points $p_i$ one by
one gives already a very good solution:
% with a much larger number $n$ of vertices than the
% systematic approach:
we choose $p_{i+1}$ from the possible grid points in the
appropriate box in such a way that the segment $p_{i+1}-p_{i}$ has the slope
as small as possible while still forming a convex angle at $p_i$.
The results in the right
column of Table~\ref{tab:exp} indicate that this algorithm is quite
competitive with the optimum solution. The running time is $O(kw)$.

\section{Grid Points in a Triangle}
\label{proof-grid}
In this section we prove Theorem~\ref{grid-in-triangle}.
We denote by $\mathbb{P} := \{\,(x,y)\mid \gcd(x,y)=1\,\}$ the set of primitive vectors in the plane.

It is known that the proportion of {primitive vectors}
among the integer vectors
in some large enough area is approximately  $1/\zeta(2) = 
6/\pi^2$~\cite[Chapters~16--18]{hw-itn-79}.
Thus, a ``large'' triangle $T$ should contain roughly
$3/\pi^2\cdot wt \approx 0.304 wt$ primitive points.
However, for very wide or very high triangles, the fraction of
primitive vectors may be different. In fact, for $t=2$, the bound
$wt/4$ is tight except for an additive slack of at most~$2$.

We will use special methods for counting primitive vectors
 when $T$ is
``very high'' (i.~e., $w$ is fixed and below some threshold and $t$ is
unbounded,
Section~\ref{HIGH}),
 when $T$ is
``very wide'' ($t$ is fixed and $w$ is unbounded,
Section~\ref{WIDE}),
and for the case when both $t$ and $w$ are large
(Section~\ref{MEDIUM}).
We use the help of the computer for the first two cases, but we use a
general bound for the last case.

\subsection{Fixed width, unbounded height}
\label{HIGH}

For a fixed value of $w$, the function $f(t) := |T \cap \mathbb{P}|$
can be analyzed explicitly. It is periodically ascending:
$$f(t+w) = f(t)+C,$$
where $C = \sum_{i=1}^{w}\phi(i)$ is the number of primitive vectors
in the triangle $(0,0)$, $(w,0)$, $(w,w)$, excluding the point
$(1,1)$.  Euler's totient function $\phi(i)$ denotes the
number of integers $1 \le j \le i$ that are relatively prime to~$i$,
or equivalently, the number of primitive vectors $(i,j)$ on the vertical
line segment from $(i,0)$ to $(i,i-1)$.

The reason for the periodic behavior
is that the unimodular shearing transformation $(x,y)\mapsto (x,y+x)$
maps the triangle
$(0,0)$, $(w,0)$, $(w,t)$,
to the triangle
$(0,0)$, $(w,w)$, $(w,w+t)$, which is equal to 
$(0,0)$, $(w,0)$, $(w,w+t)$ minus
 the triangle $(0,0)$, $(w,0)$, $(w,w)$.

 Therefore, it is sufficient to check that the ``average slope'' $C/w$
 of $f$ is bigger than $w/4$, and to check
\begin{equation}
  \label{check}
f(t)\ge tw/4  
\end{equation}
for the initial interval $2\le t \le 2+w$.  This can be done by
computer: We sort all primitive vectors $(x,y)$ with $0\le x\le w$ and
$0\le y/x \le (w+2)/w$ by their slope $y/x$. We gradually increase $t$ from $2$ to $w+2$.
The critical values of $t$ for which (\ref{check}) must be checked
explicitly are when a new primitive vector is just about to enter the
triangle.

We ran a lengthy computer check to establish (\ref{check}) for
$w=1,2,\ldots,250$ and for $2\le t \le w+2$ (and hence for all~$t$).
In addition, we checked it % (\ref{check})
for the range $w=251,252,\ldots,800$ and for $2\le t \le 250$.

\subsection{Large width}
\label{WIDE}
In this section we prove Theorem~\ref{grid-in-triangle} for
small $t$ and large~$w$.
$T$ intersects
 each horizontal line $y=i$ in a segment of length
$w-(w/t)i$.
In any set of $i$ consecutive grid points on this line, there are precisely
$\phi(i)$ primitive vectors.
We can subdivide the grid points on $y=i$ into
$\lfloor (w-(w/t)i)/i \rfloor
\ge w/i-w/t-1$
 groups of $i$ consecutive points, leading to a total of at least
$ (w/i-w/t-1)\phi(i)$ primitive vectors:
$$
  |T\cap \mathbb{P}|
 \ge 1+
\sum_{i=1}^{\lfloor t \rfloor} \left(\frac wi-\frac wt-1\right)\phi(i)
$$
For a given value of $\lfloor t \rfloor$, one can evaluate the
expression
\begin{equation}
  \label{wide-bound}
  |T\cap \mathbb{P}| 
 \ge 1+
\sum_{i=1}^{\lfloor t \rfloor} \Bigl(\frac wi-\frac wt-1\Bigr)\phi(i)
\ge 1+
\sum_{i=1}^{\lfloor t \rfloor} \Bigl(\frac wi-\frac w{\lfloor t \rfloor}-1\Bigr)\phi(i)
%=: g(w) % \lfloor t \rfloor)
\end{equation}
explicitly.
The right-hand side of this bound is a linear function $g(w)$:
\begin{equation}
\nonumber
g(w) =
\sum_{i=1}^{\lfloor t \rfloor} \Bigl(\frac wi-\frac w{\lfloor t \rfloor}-1\Bigr)\phi(i)
\end{equation}
For example, for $\lfloor t \rfloor = 130$, we have $g(w)=
w\cdot 39.514\ldots -5153$.
It follows that
$g(w)>w\cdot 131/4 > wt/4$ for $w\ge 762$.
 Performing this calculation by computer for $\lfloor t\rfloor =6,7,\ldots,130$
 establishes Theorem~\ref{grid-in-triangle} for
 $6\le t\le 130$ and $w\ge 800$.
The interval $4\le t<6$ can be split into the
ranges
 $4\le t<4.5$,
 $4.5\le t<5$,
 $5\le t<5.5$, and
 $5.5\le t<6$.
For each range, we can use the above method with a tighter bound in (\ref{wide-bound}) than
$t\ge\lfloor t\rfloor$, and the estimate goes through in the same way.

So let us consider the remaining interval $2\le t \le 4$:
For $2\le t<3$, % ($\lfloor t \rfloor=2$)
we can evaluate
$  |T\cap \mathbb{P}| $ explicitly:
\begin{equation}
  \label{eq:t4}
  |T\cap \mathbb{P}| = 1 + (w+1-\lceil\textstyle\frac wt\rceil)
+ (\lceil \frac w2\rceil-\lceil\textstyle\frac wt-\frac12\rceil),
\end{equation}
counting the primitive vectors on the lines $y=0$, $y=1$, and $y=2$,
respectively.  For $t\ge3$, the right-hand side of (\ref{eq:t4}) is still valid as a
lower bound.  We get
$$  |T\cap \mathbb{P}| \ge 1 + \textstyle\left(w+1-(\frac wt + 1)\right)
+ \left(\frac w2-( \frac wt-\frac12+1)\right)
> w (\frac 32 -\frac 2t)
$$
The last expression is $\ge wt/4$ for $2\le t \le 4$.

Thus we have proved the theorem for
 $2\le t\le 130$ and $w\ge 800$.

 \subsection{Large triangles}
\label{MEDIUM}

\begin{figure}
  \centering
\ifnarrow
  \includegraphics[width=\textwidth]{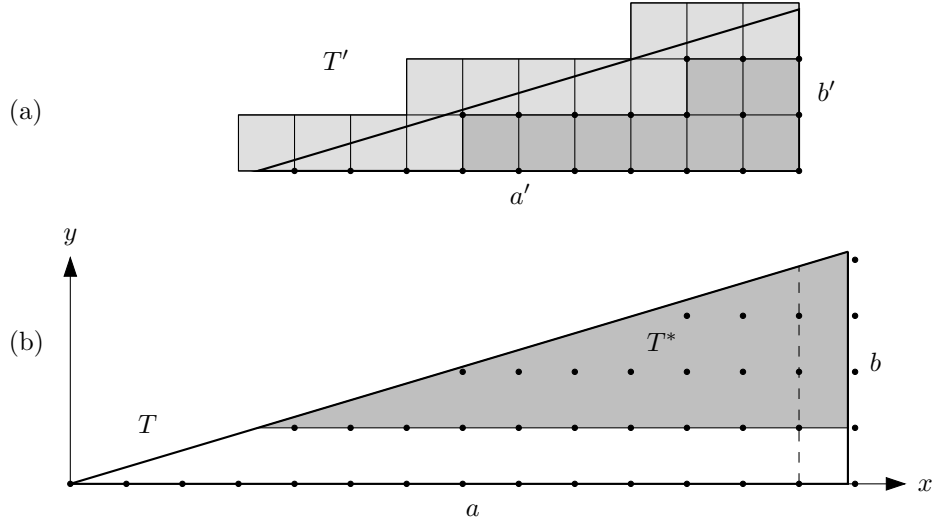}
\else
  \includegraphics{grid-lemma}
\fi
  \caption{(a) The triangle $T'$ in Lemma~\ref{right} and its covering
    by squares.
(b) The triangles $T$ and $T^*$ (shaded) in Lemma~\ref{lemma-grid-in-triangle}.}  \label{fig:tri-T}
\end{figure}

 \begin{lemma}\label{right}
   Let $T'$ be an axis-aligned right triangle of width $a'$ and
   height $b'$, whose right angle lies on a grid point.
Then
$$
\area T' 
\le
|T' \cap \mathbb{Z}^2| \le \area T' + \lfloor a' \rfloor +\lfloor b' \rfloor  +1
$$
 \end{lemma}
 \begin{proof}
 This is simple.  Suppose the right angle is at the right
 bottom corner of $T'$, see Figure~\ref{fig:tri-T}a. Each lattice
 point in $T'$ is the right bottom vertex of a unit square and these
 squares cover $T'$.
To bound the area from below, we must subtract the squares which are
not contained in $T'$.  These squares form a monotone chain along the
longest side of $T'$, and their number is $\lfloor a'\rfloor+\lfloor
b' \rfloor +1$.
 \end{proof}

\begin{lemma}
\label{lemma-grid-in-triangle}
  Let $T$ be the right triangle $(0,0)$, $(a,0)$, $(a,b)$,
with $a, b \ge 1$. Define $T^*$ as $T \cap \{\,(x,y) %\in R^2
: y\ge 1\,\}$.
Then
\begin{equation}
  \label{eq:grid-ind-triangle}
\frac {ab}2 -a -b +\frac a{2b}\leq |T^* \cap \mathbb{Z}^2| \leq \frac
{ab}2 + b -\frac a{2b}
\end{equation}
In particular,
$$
\left| \left|T^* \cap \mathbb{Z}^2\right|- \frac {ab}2 \right| \le a +b %+1
% - \frac a{2b}
$$
\end{lemma}
%If $b<1$, then $|T^* \cap \mathbb{Z}^2|=0$ and the lemma holds
%trivially.
%
%So let us assume $b\ge 1$. Then
\begin{proof}
See Figure~\ref{fig:tri-T}b.
The triangle
$T^*$ has length $a-a/b$, height $b-1$
and area $%\area T^*=
\frac 12 (b-1)(a-a/b) = ab/2-a+a/(2b)$. 
Let $T'$ denote the part of $T^*$ that lies left of the line
$x=\lfloor a \rfloor$. This triangle contains the same grid points
as $T^*$.
We assume first that $T'$ is a nonempty triangle.
The difference in areas lies in a rectangle strip of width $<1$ and height $b-1$:
$$ \area T^* - (b-1) \le \area T' \le \area T^*$$
We can apply Lemma~\ref{right} to $T'$ and obtain
$$
\begin{aligned}  
|T^* \cap \mathbb{Z}^2|
=|T' \cap \mathbb{Z}^2|
&\le \Bigl(\frac{ab}2-a+\frac a{2b}\Bigr) + \Bigl(a-\frac ab\Bigr) + (b-1) + 1,
\\
|T^* \cap \mathbb{Z}^2|
=|T' \cap \mathbb{Z}^2|
&\ge  \area T' \ge \Bigl(\frac{ab}2-a+\frac a{2b}\Bigr) - (b-1),
\end{aligned}
$$
from which the lemma follows.

\begin{figure}
  \centering
  \includegraphics{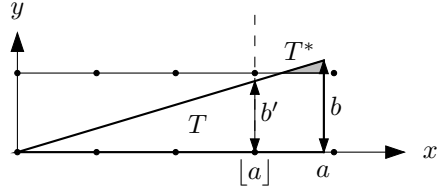}
  \caption{If $T^*$ (shaded) contains no grid points, the triangle $T'$
  does not exist.}
 \label{fig:grid-lemma-small-case}
\end{figure}

The triangle $T'$ may not exist, as in
Figure~\ref{fig:grid-lemma-small-case}.
In this case, $T^* \cap \mathbb{Z}^2 = \emptyset$.
  Instead of arguing why the above
derivation is valid also for this case, we establish the inequalities
directly.  Let $b' \ge b-b/a$ denote the vertical extent of $T$ at $x=\lfloor
a\rfloor$.  Then the fact that $T'$ is empty is equivalent to $b'<1$.

Then, from $1\ge b' \ge b-b/a$ we conclude that $ab<a+b$.
It follows that the lower bound in
(\ref{eq:grid-ind-triangle})
is at most~0:
$$
\frac{ab}2 +\frac{a}{2b} -a-b \le \frac{a+b}2 +\frac{a}{2} -a-b \le
0$$
The claimed upper bound in (\ref{eq:grid-ind-triangle}) is always
nonnegative, by the assumption $b\ge1$.
\end{proof}

The number of primitive vectors can be estimated by an inclusion-exclusion
formula, taking into account vectors which are multiples of single primes
$2,3,5,7,\ldots$, vectors which are jointly multiples of two primes, of three
primes, and so on, see \cite[Chapters~16--18]{hw-itn-79}:
\begin{equation}
  \label{eq:inclusion-exclusion}
|T\cap \mathbb{P}| 
= 1+|T^*\cap \mathbb{P}| 
= 1+\sum_{i=1}^S \mu(i)\cdot
\left|(\tfrac 1i\cdot T^*)\cap \mathbb{Z}^2\right|
= 1+\sum_{i=1}^S \mu(i)\cdot
\left|({\tfrac 1i} T)^*\cap \mathbb{Z}^2\right|
\end{equation}
Here, $\mu(i)$ is the M\"obius function: $\mu(i)=(-1)^k$ if $i$ is the
product of $k$ distinct primes and $\mu(i)=0$ otherwise.  It is known
that $\sum_{i=1}^\infty \frac {\mu(i)}{i^2} =1/\zeta(2) = 6/\pi^2$,
leading to the fact mentioned above that a fraction of approximately
$6/\pi^2$ of the grid points in a large area are primitive vectors.

Our sum in (\ref{eq:inclusion-exclusion}) goes to $i=\infty$, but for $i>w$ or $i>t$,
the set
$({\textstyle \frac 1i} T)^*\cap \mathbb{Z}^2$ is empty.
Therefore, the formula is valid for $S := \min\{w,\lfloor
t\rfloor\}$.
We apply Lemma~\ref{lemma-grid-in-triangle} and obtain
\begin{align*}
|T\cap \mathbb{P}| %&
= 1+\sum_{i=1}^S \mu(i)\cdot
\left|({\textstyle \frac 1i} T)^*\cap \mathbb{Z}^2\right|
%\\
&
\ge \frac {wt}2 \sum_{i=1}^S \frac {\mu(i)}{i^2}
- \sum_{i=1}^S  \frac {w+t}i %- \frac w{2t} - 1
\\&
\ge \frac {wt}2 \left(\frac 6{\pi^2}-\frac 1S \right)
- H_S (w+t), %+ \frac {Sw}{2t} 
%-S
\end{align*}
where $H_S=1+1/2+1/3+\cdots +1/S$ is the harmonic number. The last inequality
comes from bounding the remainder
$\sum_{i=S+1}^\infty \mu(i)/i^2 \le\sum_{i=S+1}^\infty 1/i^2<1/S$
of the infinite series, whose value is $6/\pi^2.$

We distinguish the two cases for~$S$:
Case 1: $w\le t$, and $S=w$. Then
\begin{align*}
|T\cap \mathbb{P}| &
 \ge \frac {wt}2 \left(\frac 6{\pi^2}-\frac 1w \right)
 - H_w (2t) % -t % + \frac {Sw}{2t} -w
%\\ &
=wt \left(\frac 3{\pi^2}-\frac 1{2w} -\frac {2H_w}w\right)
\end{align*}
Case 2: $w\ge t$, and $S=\lfloor t\rfloor$.
% \begin{align}
% |T\cap \mathbb{P}| &
%  \ge \frac {wt}2 \left(\frac 6{\pi^2}-\frac 1{\lfloor t\rfloor} \right)
%  - H_{\lfloor t\rfloor} (w+t)  + \frac {\lfloor t\rfloor w}{2t} 
% - t % \lfloor t\rfloor
% \nonumber \\ &
% \ge wt \left(\frac 3{\pi^2}-\frac 1w -  H_{\lfloor t\rfloor} \left(\frac
%     1w+\frac 1t\right)  - \left[\frac 1{2(t-1)}-\frac{t-1}{2t^2}\right]\right)
% \nonumber\\ &
% = wt \left(\frac 3{\pi^2}-\frac 1w -  H_{\lfloor t\rfloor} \left(\frac
%     1w+\frac 1t\right)  -\frac1{2t^2(t-1)}\right)
% \nonumber
% \end{align}
\begin{align}
|T\cap \mathbb{P}| &
 \ge \frac {wt}2 \left(\frac 6{\pi^2}-\frac 1{\lfloor t\rfloor} \right)
 - H_{\lfloor t\rfloor} (w+t) % - t % \lfloor t\rfloor
\nonumber \\ &
\ge wt \left(\frac 3{\pi^2} - \frac 1{2(t-1)} % -\frac 1w 
-  H_{\lfloor t\rfloor} \left(\frac
    1t+\frac 1w\right) 
\right)
\label{w-large}
\end{align}

Combining the two cases and setting $n := \min \{w,t\}$ gives
$$|T\cap \mathbb{P}| \ge 
 wt \left(\frac 3{\pi^2} %-\frac 1{n-1} 
-\frac 1{2(n-1)} - \frac{2 H_{\lfloor n\rfloor}}n
\right)
$$
Using the estimate $ H_i \le \gamma + \ln (i+1)$
with Euler's constant $\gamma \approx 0.57721$,
it can be checked that this factor is bigger than $1/4$ for $n\ge
250$, % even for 233
thus proving the theorem for $w,t\ge 250$.

On the other hand, the factor in (\ref{w-large}) is 
bigger than $1/4$ for $w\ge 800$ and $130\le t \le 250$,
proving the theorem also for this range.

\paragraph{Wrap-up.}
The proof of Theorem~\ref{grid-in-triangle} is now complete.
On a high level, we distinguish three ranges for $w$:
 $1\le w \le 250$,  $251\le w \le 800$, and $w\ge 800$.
 \begin{itemize}
 \item 
Range 1:
For $1\le w \le 250$, the theorem has been established in
Section~\ref{HIGH}.

\item Range 2: $251\le w \le 800$.  For $251\le w \le 800$ and $1\le t
  \le 250$, the theorem has been established in Section~\ref{HIGH} as
  well.  For $251\le w \le 800$ and $t \ge 250$, it has been proved in
  Section~\ref{MEDIUM}.
\item 
Range 3:
Finally, for $w\ge 800$, there is a division into three cases:
Section~\ref{WIDE} takes care of the range $2\le t \le 130$.
Section~\ref{MEDIUM} proves the bound separately for the ranges
$130\le t <250$ and $t \ge 250$. 
\qed
 \end{itemize}

\section{Conclusion}
\label{sec:conclusion}

In practice, the algorithm behaves much better than indicated by the
rough worst-case bounds that we have proved.
We have not attempted to optimize the constants in the proof.
For example, if we don't take a $7\times 7$ subgrid but an $11\times
11$ subgrid,
and with a more specialized treatment of the outer face,
the permissible amount of perturbation in
Lemma~\ref{small-perturb} increases from $1/30$ to $1/9$,
% THIS IS TRUE: Different perturbation disks from one point are
% even separated by a horizontal line.
but it would make the pictures of the rough perturbation harder to draw.

Bonichon, Felsner, and Mosbah~\cite{bfm-cd3cp-05} have used a technique of
eliminating edges from the drawing that can later be inserted in order to
reduce the necessary grid size for (non-strictly) convex drawings. This
technique can also be applied in our case: remove interior edges as long
as the graph remains three-connected. These edges can be easily reinserted in
the end, after all faces are strictly convex.  (For non-strictly convex
drawings in~\cite{bfm-cd3cp-05}, the selection of removable edges and their reinsertion is actually a
more complicated issue.)  This technique might be useful in practice for
reducing the grid size.
% constant factors

\paragraph{Lower Bounds.}
The only known lower bound comes from the fact that a single convex
$n$-gon on the integer grid needs $\Omega(n^3)$ area, see B\'ar\'any
and Tokushige~\cite{bt-macln-04}, or Acketa and {\v
  Z}uni{\'c}~\cite{az-mnecd-82a,az-mnecd-95} for the easier case of a
\emph{square} grid.  To achieve this area for an $n$-gon, one has to
draw it in a quite round shape. In contrast, the faces that are
produced in our algorithm have a very restricted shape: when viewed
from a distance, the look like the triangles, quadrilaterals,
pentagons, or hexagons of the $n\times n$ grid drawing from which they
were derived.  To reduce the area requirement below $O(n^4)$ one has
to come up with a new approach that also produces faces with a
``rounder'' shape.

Our bounds are however, optimal within the restricted class of
algorithms that start with a Schnyder drawing or an arbitrary
non-strictly convex drawing on an $O(n)\times O(n)$ grid and try to
make it strictly convex by \emph{local perturbations} only. %, as our algorithm.
Consider the case where $n-1$ vertices lie on the outer face,
connected to a central vertex in the middle.  The Schnyder drawing
will place these vertices on the enclosing triangle, and at least
$n/3$ vertices will lie on a common line. They have to be perturbed
into convex position, as in Figures~\ref{fig:fine-perturb-overview} or
\ref{fig:grid}.

Let us focus on the standard situation when we want to perturb $n$ equidistant
vertices on a line, at distance~1 from each other.  The $n-1$ edge vectors
$p_{i+1}-p_i$ lie in a $2w\times 2h$ box; they must be non-parallel, and
in particular, they must be \emph{distinct}. If $\Delta y$ is the average absolute
vertical increment of these vectors, it follows that $\Delta y=\Omega(n/w)$,
and the total necessary height $h$ of the boxes is $\Omega(n(\Delta
y))=\Omega(n^2/w)$.  Therefore,
% $hw=\Omega(n^2)$, and
the total necessary area is $\Omega(hwn^2)=\Omega(n^4)$.

The argument can be extended to the case when only $\Omega(n)$
selected grid vertices on a line of length $O(n)$ have to be
perturbed. It can also be shown that our bounds in terms of $k$ are
optimal in this setting. The worst case occurs when there is a line of
length $n$ with $\Omega(k)$ consecutive grid points in the middle and
two vertices at the extremes.

\paragraph{Extensions.}
The class of three-connected graphs is not the most general class of graphs
which allow strictly convex embeddings.
The simplest example of this is a single cycle.  A planar graph, with a specified face
cycle $C$ as the outer boundary,
has a strictly convex embedding if and only if it is % \emph{internally}
three-connected to the boundary, i.~e., if every interior vertex (not on $C$)
has three vertex-disjoint paths to the boundary cycle.  Equivalently, the
graph becomes three-connected after adding a new vertex and connecting it to
every vertex of~$C$.
These graphs cannot be treated directly by our approach, since the
Schnyder embedding method of Felsner~\cite{f-cdpgo-01} does not apply.
Partitioning the graph into three-connected components and putting
them together at the end might work.

\paragraph{Acknowledgements.}
We thank the referees for helpful remarks which have lead to many
clarifications in the presentation.
Imre B\'ar\'any was partially supported by Hungarian National
  Science Foundation Grants No.\ T~037846 and T~046246.

{%\small
\bibliographystyle{amsplain}
%\bibliography{drawing-strictly,../geombib/geom}
\bibliography{drawing-strictly,geom}

}

\end{document}